\documentclass[twocolumn,times,twocolappendix]{aastex631}

\usepackage{amsmath,amstext}
\usepackage{CJK}
\usepackage[T1]{fontenc}
\usepackage{newtxtext,newtxmath}
\usepackage[figure,figure*]{hypcap}
\usepackage{booktabs}
\usepackage{cleveref}
\usepackage{paralist}
\usepackage{ulem}
\usepackage[colorlinks=true]{hyperref}

\newcommand{\fb}{f_\mathrm{b}}
\newcommand{\gq}{\gamma_q}

\newcommand{\amf}{\alpha_\mathrm{mf}}
\let\oldmid\mid
\renewcommand{\mid}{\!\oldmid\!}
\newcommand{\mass}{\mathcal{M}}

\newcommand{\BPRP}{ {G_\mathrm{BP}-G_\mathrm{RP} }}

\newcommand{\mix}{\mathrm{mix}}
\newcommand{\fds}{\mathrm{fs}}
\newcommand{\BF}{\mathrm{BF}}

\begin{document}
\begin{CJK*}{UTF8}{gbsn}

\title{Validating Open Cluster Candidates with Photometric Bayesian Evidence}

\correspondingauthor{Lu Li}
\email{lilu@shao.ac.cn}

\author[0000-0002-0880-3380]{Lu Li (李璐)}
\affil{Shanghai Astronomical Observatory, Chinese Academy of Sciences, 80 Nandan Road, Shanghai 200030, China.}

\author[0000-0001-7890-4964]{Zhaozhou Li (李昭洲)}
\affil{School of Astronomy and Space Science, Nanjing University, Nanjing, Jiangsu 210093, China}
\affil{Key Laboratory of Modern Astronomy and Astrophysics, Nanjing University, Ministry of Education, Nanjing 210093, China}
\affil{Centre for Astrophysics and Planetary Science, Racah Institute of Physics, The Hebrew University, Jerusalem, 91904, Israel}

\author[0000-0001-8611-2465]{Zhengyi Shao (邵正义)}
\affil{Shanghai Astronomical Observatory, Chinese Academy of Sciences, 80 Nandan Road, Shanghai 200030, China.}
\affil{Key Lab for Astrophysics, Shanghai 200234, China}

\begin{abstract}

The thousands of open cluster (OC) candidates identified by the Gaia mission are significantly contaminated by false positives from field star fluctuations, posing a major validation challenge. Based on the Mixture Model for OCs (MiMO), we present a Bayesian framework for validating OC candidates in the color--magnitude diagram. The method compares the Bayesian evidence of two competing models: a single stellar population with field contamination versus a pure field population. Their ratio, the Bayes factor (BF), quantifies the statistical support for cluster existence. Tests on confirmed clusters and random fields show that a threshold of BF > 100 effectively distinguishes genuine clusters from chance field overdensities. This approach provides a robust, quantitative tool for OC validation and catalog refinement. The framework is extendable to multi-dimensional validation incorporating kinematics and is broadly applicable to other resolved stellar systems, including candidate moving groups, stellar streams, and dwarf satellites.

\end{abstract}

\keywords{Open star clusters (1160), Hertzsprung Russell diagram (725), Mixture model (1932), Bayesian statistics (1900)}

\section{Introduction}

Open clusters (OCs) are key tracers of star formation and Galactic evolution. As nearly coeval stellar populations, they provide critical constraints on star formation and evolution, cluster dynamics, and structure and chemistry of the Milky Way disk.

The \textit{Gaia} mission has dramatically expanded the known OC population. While initial discoveries in the Gaia era were sometimes made by visually inspecting astrometric space \citep{sim207NEWOPEN2019}, the large volume of data quickly necessitated more automated approaches. Unsupervised clustering algorithms in the five-dimensional astrometric space (sky coordinates, proper motions, and parallax)  such as Gaussian Mixture Model \citep{cantat-gaudinGaiaDR2Unravels2019}, DBSCAN \citep{castro-ginardNewMethodUnveiling2018a,castro-ginardHuntingOpenClusters2022} and HDBSCAN \citep{hunt2023} have been particularly effective, enabling large-scale, all-sky surveys that have significantly increased the OC census.

However, a critical challenge arising from these automated searches is the prevalence of false positives. Not all kinematic and spatial overdensities correspond to a genuine Single Stellar Population (SSP) of common origin; many are simply chance alignments or random fluctuations of field stars. Such contaminated catalogs directly impact the reliability of studies, not only of OC properties but also of stellar evolution and Galactic archaeology that rely on OCs as tracers.
Consequently, robust validation of these candidates has become a crucial step in modern OC research.

A common way to validate such candidates is by inspecting their color--magnitude diagram (CMD) for a narrow, isochrone-like sequence, which indicates a physical SSP. This assessment is often performed visually or semi-empirically. 
Modern quantitative methods include training machine learning classifiers, such as artificial neural networks, to recognize CMD patterns of real clusters \citep{cantat-gaudinGaiaDR2Unravels2019,hunt2023}, or assigning quality scores based on astrometric density and empirical photometric likelihood to quantify how distinct the member stars are from a random field population \citep{perrenUnifiedClusterCatalogue2023}.

However, validation in the CMD can be delicate and even misleading. While a narrow isochrone clearly confirms a prominent cluster, the challenge lies with ambiguous candidates with a less obvious main sequence. For these kinematically selected groups, the observed scatter in the CMD makes it difficult to distinguish whether the pattern represents a true cluster broadened by observational errors and differential reddening, or just a collection of unrelated field stars. This ambiguity arises because even a sample of field stars can form a relatively narrow main sequence if drawn from a small Galactic volume, as exemplified by the original Hertzsprung-Russell diagram, which was first constructed from field stars in the solar neighborhood. Therefore, a more robust and interpretable validation requires demonstrating the statistical necessity of an additional SSP component (including single stars and unresolved binaries) on top of the expected field background. This directly motivates the use of a mixture model to identify false positives.

In this work, we present a principled solution using rigorous Bayesian model comparison, implemented through our Mixture Model for OCs (MiMO; \citealt{liModelingUnresolvedBinaries2020,liMiMOMixtureModel2022}). MiMO models the CMD distribution of stars as a mixture of a single stellar population (including unresolved binaries) and a non-parametric field population. While primarily used to find best-fit cluster parameters (e.g., isochrone and stellar mass function), MiMO also computes the Bayesian evidence of this model. This evidence is then compared to that of a pure field-star model, and their ratio, the Bayes factor, quantifies the statistical support for the presence of an SSP, the indicator of a real OC.

We apply this method to both confirmed clusters and random field samples, demonstrating that the Bayes factor effectively distinguishes real OCs from spurious overdensities and offers a robust statistical criterion for cluster validation in the era of large-scale surveys. 
Although we focus on OCs, this photometric evidence is broadly applicable for validating any candidate coeval population (e.g., moving groups or stellar streams) against a field background.

\section{Method: Bayesian Evidence} \label{sec:method}

We compute the Bayesian evidence using the MiMO framework. A brief summary is provided in the Appendix, while full methodological details, including validation with mock samples and numerical implementation, are given in \citealt{liMiMOMixtureModel2022}.

The observed CMD distribution of a stellar sample is modeled as a mixture of cluster members, $\phi_\mathrm{cl}$, and field stars, $\phi_\mathrm{fs}$,
\begin{align}
\label{eqn:phitot}
    \phi_\mix(m,c \mid \Theta) = (1-f_\mathrm{fs}) \phi_\mathrm{cl}(m,c \mid \Theta) + f_\mathrm{fs}\phi_\mathrm{fs}(m,c),
\end{align}
where $(m,c)$ denote the apparent magnitude and color, and $\Theta$ is the set of model parameters, including the isochrone parameters (age, metallicity, distance modulus, and extinction), the stellar mass function slope, the binary fraction, and the fraction of field stars in the sample, $f_\mathrm{fs}$.
Each component is normalized such that $\int\!\phi(m,c) dmdc=1$.

Cluster members are modeled as a mixture of single stars and unresolved binaries, with distributions shaped by the isochrone (age, metallicity, distance, extinction), stellar mass function, binary fraction, and binary mass-ratio distribution. Specifically, we adopt PARSEC isochrones \citep{bressanPARSECStellarTracks2012} with the Gaia EDR3 photometric system \citep{rielloGaiaEarlyData2021}, and the variable extinction model YBC \citep{chenYBCStellarBolometric2019}. The field population is modeled empirically using adaptive kernel density estimation from an auxiliary sample of neighboring field stars for each cluster, assuming they represent the same population as the field stars within the cluster region.

Given a sample of $N$ stars, $D = \{m_i, c_i\}_{i=1}^{N}$, the likelihood under the mixture model is
\begin{equation}\label{eq:lhtot}
    \mathcal{L}_\mix(D \mid \Theta) = \prod\nolimits_{i=1}^{N}\ \phi_\mix(m_i, c_i \mid \Theta).
\end{equation}
The posterior distribution of the model parameters follows from Bayes' theorem,
\begin{equation}\label{eq:pdf}
    P_\mix (\Theta \mid D) = 
    	\frac{\mathcal{L}_\mix(D \mid \Theta) \pi(\Theta)}{P_\mix(D)},
\end{equation}
where $\pi(\Theta)$ is the prior. In this work, we adopt flat priors (see Table 1 in \citealt{liMiMOMixtureModel2022}).
The normalization term,
\begin{equation}\label{eq:ev_mix}
    P_\mix (D) = \int_\Theta {\mathcal{L}_\mix(D \mid \Theta) \pi(\Theta)} d\Theta
\end{equation}
is the \emph{Bayesian evidence} of the mixture model (see, e.g., \citealt{trottaBayesSkyBayesian2008}). It quantifies the average likelihood under the prior and enables quantitative comparison between models.

To assess the significance of the cluster component, we compare this to a competing model in which all stars are field stars,
\begin{equation}\label{eq:ev_fs}
    P_\fds (D) = \mathcal{L}_\fds(D) = \prod\nolimits_{i=1}^{N}\ \phi_\fds(m_i, c_i),
\end{equation}
which has no free parameters and hence its evidence is simply the likelihood.
The Bayes factor,
\begin{equation}\label{eq:bayes_fac}
    \BF \equiv \frac{P_\mix(D)}{P_\fds(D)},
\end{equation}
quantifies the statistical support for the presence of an SSP in the CMD. Unlike classical goodness-of-fit metrics, the Bayes factor incorporates the full posterior volume and observational uncertainties, and penalizes model complexity.

We compute $P_\mix(D)$ using the nested sampling algorithm \citep{skillingNestedSampling2004}, as implemented in the \texttt{dynesty} package \citep{speagleDynestyDynamicNested2020}.\footnote{\url{https://github.com/joshspeagle/dynesty}} This process also yields weighted posterior samples of $\Theta$, simultaneously enabling parameter inference.

\section{Bayes factor of random field and real OCs} \label{sec:bf_dist} 

\begin{figure*}\label{fig:cmd}
    \centering
    \includegraphics[width=\linewidth]{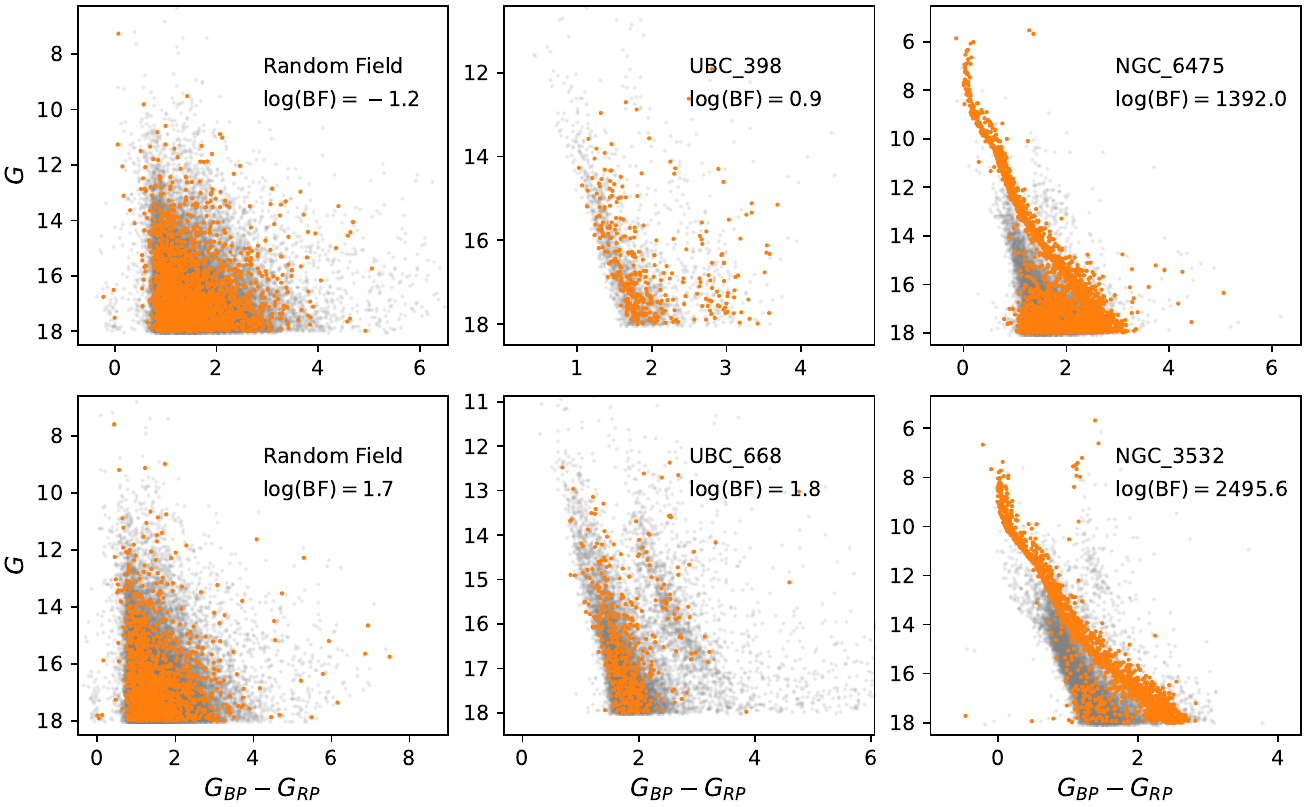}
    \caption{Example CMDs for different types of targets. Left: Random field regions ($\log_{10}\mathrm{BF} < 0$). Middle: Ambiguous candidates ($\log_{10}\mathrm{BF} \sim 1$). Right: Confirmed OCs with strong evidence ($\log_{10}\mathrm{BF} > 3$). BF denotes the \emph{Bayes factor} comparing a mixture model of SSP+field to a pure field-star model.
    }
    \label{fig:cmd_examples}
\end{figure*}

In this section, we evaluate the Bayes factor's ability to distinguish real open clusters from false positive candidates arising from pure random field star fluctuations. 
In principle, a sample composed entirely of field stars should yield a Bayes factor typically less than unity, as the field-only model provides a better explanation. Conversely, a sample containing a genuine single stellar population should produce a large Bayes factor, indicating that the mixture model is statistically preferred.

To test this, we constructed 600 random field samples by drawing subsets of $N$ stars from the full sky Gaia DR3 catalog with $G < 18$. The stars in each sample have unrelated distances and evolutionary stages. The number of stars in each subset, $N$, ranged from 20 to 3000 to span the varying richness of real cluster candidates, although we found that the value of $N$ does not significantly affect the Bayes factors.
For each random sample, we computed the Bayes factor by comparing the evidence of the mixture model against the field-only model.

The results confirm our expectations. As shown in Figure \ref{fig:cmd}, random field samples (left panels) consistently yield $\log_{10}\text{BF} \lesssim 0$, indicating that the field-only model is favored. For comparison, we analyzed several known OC candidates from \citet{cantat-gaudinGaiaDR2Unravels2019} and \citet{diasUpdatedParameters17432021}. 
Visually ambiguous candidates yield $\log_{10}\text{BF} \sim 1$ (middle panels), which quantitatively confirms that their CMD distribution is almost indistinguishable from that of the field. In contrast, well-defined, confirmed clusters with main sequence distinguished from the field have strong statistical evidence and large Bayes factors (right panels). These examples demonstrate the utility of the Bayes factor for robustly classifying cluster candidates.

 \begin{figure}[!tbp]\label{fig:bf_hist}
  \centering
  \includegraphics[width=1\columnwidth]{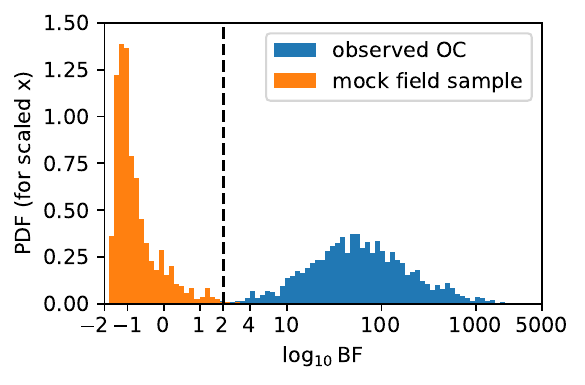}
  \caption{Distributions of Bayes factor (BF) for random field samples (orange) and confirmed OCs (blue). To accommodate the wide dynamical range of $\log_{10}\mathrm{BF}$, the $x$-axis adopts an arcsinh scale, which is linear near zero and logarithmic at large values. The probability density on the $y$-axis is computed for $\mathrm{arcsinh}(\log_{10}\mathrm{BF})$ accordingly. A threshold of $\log_{10}\mathrm{BF} \simeq 2$ effectively separates real clusters from field samples.
  }
  \label{fig:dlogz_hist}
\end{figure}

Figure~\ref{fig:bf_hist} shows the distribution of $\log_{10}\mathrm{BF}$ for 600 random field samples compared to that of 1232 confirmed OCs from the MiMO catalog (L. Li et al. 2025, in press). The two populations are well-separated: confirmed OCs consistently show large Bayes factors, while random fields cluster at lower values. Based on this separation, we adopt a conservative and robust threshold of $\log_{10}\mathrm{BF} > 2$, corresponding to the cluster+field model being at least 100 times more probable than the field-only model.

This threshold provides a clean separation between the two samples. Only two of the 600 random field samples (0.4\%) exceed this value, while only one (Gulliver\_52, $\log_{10}\mathrm{BF} = 1.62$) of the 1232 confirmed OCs falls below it. Any candidate with a Bayes factor below this threshold is thus considered to lack sufficient statistical evidence to be classified as a genuine cluster.

It is important to note that this is a statistical, not a perfect, separation. A small fraction of random field alignments may produce a $\log_{10}\mathrm{BF} > 2$ by chance, and some sparse real clusters may fall below it. Furthermore, the confirmed OCs used for this test were visually vetted, introducing a selection bias that favors clusters with strong evidence. Therefore, a low Bayes factor does not definitively rule out the existence of a cluster; rather, it indicates that the photometric evidence alone is not strong enough to confirm it. We recommend reporting the Bayes factor itself as the primary output, allowing for nuanced interpretations of ambiguous cases.


\section{Discussion} \label{sec:discussion}

\subsection{Caution in field model construction}

The interpretation of the Bayes factor depends critically on one key assumption: the field-star model accurately represents the true contamination in the candidate region. Because the Bayes factor quantifies how much better the mixture model explains the data compared to the field-only model, any systematic mismatch in the field component can disfavor the field-only model. For example, if the field model is constructed from stars with a distance distribution or sky region different from those in the candidate sample, its CMD may not reflect the local field contamination, potentially leading to an overestimated Bayes factor.

 \begin{figure}[!tbp]\label{fig:bf_control}
  \centering
  \includegraphics[width=1\columnwidth]{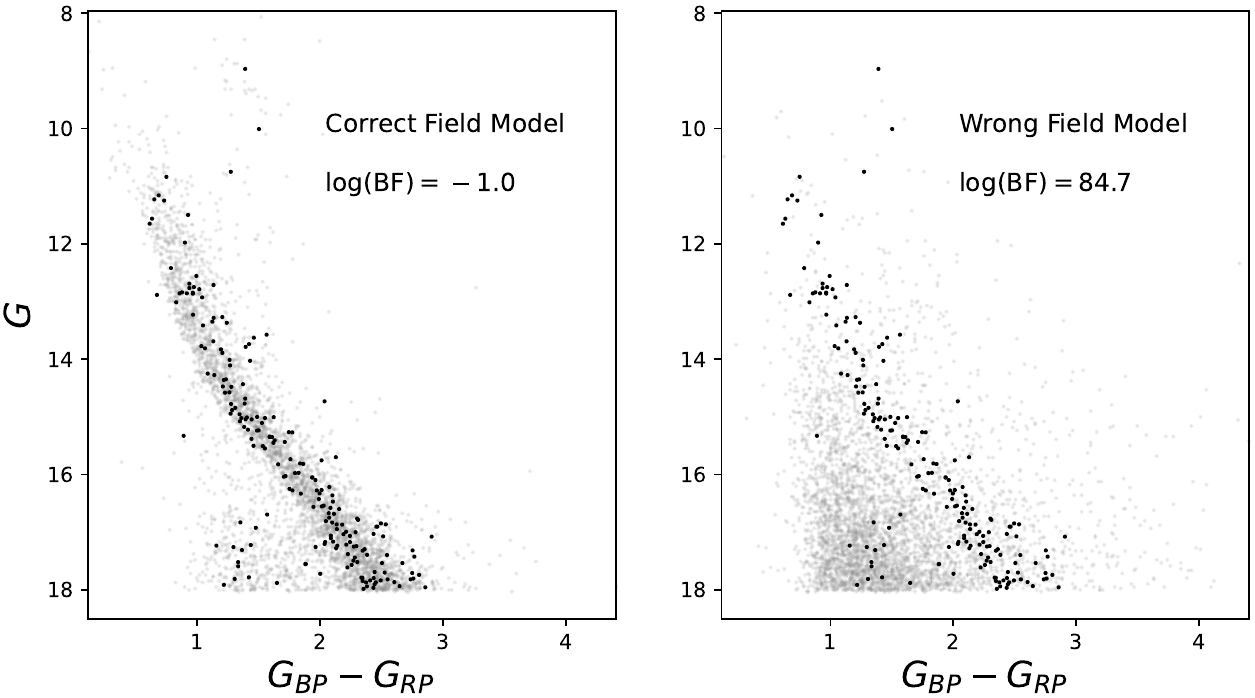}
  \caption{CMD examples illustrating the impact of field-star model mismatch. In both panels, black points represent 200 field stars within 100 pc. Left: using a representative field model built from stars in the same distance range (gray points). Right: using a mismatched field model built from a broader distance range ($d < 2$ kpc).
  }
\end{figure}

Figure~\ref{fig:bf_control} illustrates this effect with a controlled example. In both panels, the black points represent 200 field stars within 100~pc, forming a broad main sequence in the CMD that could be misidentified as a real cluster under visual inspection. When the field model is constructed using stars from a comparable distance range, the Bayes factor correctly identifies the sample as consistent with a pure field population ($\log_{10}\mathrm{BF} = -1.0$). In contrast, when the field model is built from a much broader distance range ($d < 2$~kpc), the mismatch in CMD morphology results in a spuriously high Bayes factor ($\log_{10}\mathrm{BF} = 84.9$). This example underscores the importance of constructing representative field models.

In practical applications, we recommend building empirical field-star models using stars from the vicinity of each candidate on the sky and selected to have similar parallax distributions. This ensures a fair comparison between models and a more reliable Bayes factor estimate.

\subsection{Insensitivity to cluster field contamination}

A key advantage of our Bayesian framework is its robustness against field contamination within a candidate cluster. 
Because MiMO explicitly accounts for the field fraction, the ratio structure of the Bayes factor naturally cancels the contribution from this shared field component when comparing a SSP+field model to a field-only model. 
Therefore, the mixture model isolates the statistical evidence for the excess signal from a single stellar population even under high contamination.

 \begin{figure}[!tbp]\label{fig:f_fs_vs_BF}
  \centering
  \includegraphics[width=1\columnwidth]{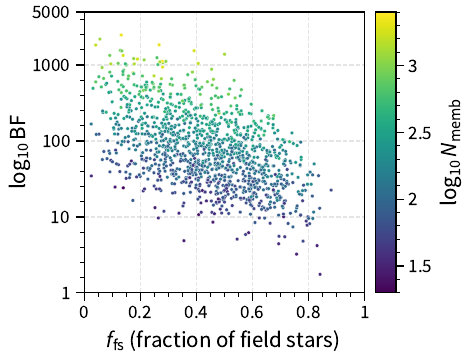}
  \caption{Relation between the Bayes factor (BF) and the field star fraction ($f_\mathrm{fs}$) for the 1232 confirmed OCs from the MiMO catalog. Points are color-coded by the number of inferred member stars ($N_\mathrm{memb}$).
  }
\end{figure}

Figure~\ref{fig:f_fs_vs_BF} demonstrates this robustness using the 1232 confirmed OCs from the MiMO catalog (L. Li et al. 2025, in press).
A strong correlation appears with the number of inferred member stars ($N_{\rm memb} = N_\mathrm{total} [1-f_{\rm fs}]$): richer clusters yield stronger evidence as expected.
In contrast, at a given $N_{\rm memb}$, the Bayes factor shows only a very weak dependence on $f_{\rm fs}$, even for highly contaminated samples ($f_{\rm fs} > 0.8$). Thus, our validation metric remains a reliable indicator of the intrinsic cluster signal regardless of sample purity.

\subsection{Bayes factor is not a goodness-of-fit metric}

It is important to recognize that the Bayes factor and the goodness-of-fit metric are distinct concepts that address different questions. The Bayes factor serves as a tool for \textit{model selection}, evaluating the overall plausibility of a model across its entire parameter space. In contrast, goodness-of-fit assesses how well a single best-fit parameter set reproduces the observed data.

A high Bayes factor therefore indicates that the presence of a cluster component is statistically necessary, but it does not imply that the resulting best-fit model provides an accurate or detailed description of the data. Conversely, a low Bayes factor signifies insufficient evidence for the cluster model, rendering any subsequent parameter inference unreliable or meaningless.

\section{Conclusion} \label{sec:conclusion}

We have presented a robust Bayesian framework, implemented with our Mixture Model for OCs (MiMO), to validate open cluster (OC) candidates by quantifying the statistical necessity of a Single Stellar Population (SSP) in their color--magnitude diagrams. By comparing the Bayesian evidence of an SSP+field mixture model against a pure field-star model, we derive a Bayes factor (BF) that serves as a direct, physical measure of whether a candidate is a genuine cluster.

Our analysis of both confirmed OCs and random field samples demonstrates that a conservative threshold of $\log_{10}\mathrm{BF} > 2$ (i.e., the SSP+field model is 100 times more probable than the field-only model) effectively distinguishes genuine, coeval populations from false positives. This photometric evidence is robust against high levels of field contamination and provides a quantitative, reproducible validation metric, a critical tool for purifying the thousands of kinematically selected OC candidates from large-scale surveys.

The impact of this framework extends beyond OCs. The general methodology of comparing a population model against a field background is readily applicable to other resolved stellar systems, including moving groups, stellar streams, and dwarf satellites of the Milky Way, where an SSP can be replaced by multiple populations following an assumed star formation history.
This validation framework can also be extended to incorporate mixture models in space and kinematics, enabling multi-dimensional validation of candidates.

The source code of MiMO is publicly available on GitHub\footnote{\url{https://github.com/luly42/mimo}}.
The MiMO catalog of 1232 OCs used in this work is available on the National Astronomical Data Center China-VO Paper-Data service (DOI: \href{https://doi.org/10.12149/101693}{10.12149/101693}), along with a copy of the code and the model isochrone files.

\section*{Acknowledgments}
We thanks Prof.~Chao Liu and Prof. Song Huang for helpful discussions and suggestions. This work is supported by the National Natural Science Foundation of China (NSFC) under grant No. 12303026 and 12273091; the Science and Technology Commission of Shanghai Municipality (Grant No. 22dz1202400); the science research grants from the China Manned Space Project with No. CMS-CSST-2021-A08. This work was also sponsored by the Young Data Scientist Project of the National Astronomical Data Center and the Program of Shanghai Academic/Technology Research Leader. 
ZZL acknowledges the Marie Skłodowska-Curie Actions Fellowship under the Horizon Europe programme (101109759, ``CuspCore'').

\vspace{5mm}

\software{Astropy \citep{astropycollaborationAstropyCommunityPython2013,astropycollaborationAstropyProjectBuilding2018,astropycollaborationAstropyProjectSustaining2022}, Dynesty, MiMO\citep{liMiMOMixtureModel2022}}

\appendix

\section{MiMO Framework in a Nutshell}\label{sec:appendix_mimo}

We briefly recap MiMO below
and refer readers to \citet{liMiMOMixtureModel2022} for more details, including the justification of the model choices, the validation and performance benchmark with mock samples, and the numerical implementation.
We emphasize that MiMO, as a general framework, can naturally apply to
any other choices for the stellar evolution model, photometric bands, extinction model, functional forms of the MF, or binary distribution.

\subsection{Mixture model} \label{sec:MM}

We consider the probability density distribution of stars in the CMD, $\phi(m,c\mid\Theta)$, in terms of the apparent magnitude $m$ and color $c$, characterized by a set of model parameters, $\Theta$ (see Table 1 in \citealt{liMiMOMixtureModel2022}).
The sample of stars follow a mixture distribution of cluster members, $\phi_\mathrm{cl}$, and field stars, $\phi_\mathrm{fs}$,
\begin{align}
\label{eqn:phitot}
    \phi_\mix(m,c \mid \Theta) = (1-f_\mathrm{fs}) \phi_\mathrm{cl}(m,c \mid \Theta) + f_\mathrm{fs}\phi_\mathrm{fs}(m,c),
\end{align}
where $f_\mathrm{fs}$ is the fraction of field stars in the sample.

In addition to cluster-level parameter inference, the Bayesian formalism can also provide the membership probability and stellar parameters (mass and binary mass ratio) for individual stars
(\citealt{liuPhotometricDeterminationUnresolved2025b}).

\subsection{Model of member stars} \label{sec:MM:cl}

The member stars follow a mixture distribution of single stars, $\phi_\mathrm{s}$, and unresolved binaries, $\phi_\mathrm{b}$,
\begin{align}\label{eq:phi-cl}
  \phi_\mathrm{cl}(m,c \mid \Theta) = (1-f_{\rm b}) \phi_\mathrm{s}(m,c \mid \Theta) + f_{\rm b} \phi_{\rm b}(m,c \mid \Theta)
\end{align}
where $\fb$ is the fraction of binary stars.

Given an isochrone specified by $\Theta$, 
the mass of a single star $\mass$ determines its location in the CMD.
The observed location is further blurred by observational uncertainties, $\sigma_{m}$ and $\sigma_{c}$ (which can be different for different stars), as
\begin{equation}\label{eq:pmc}
	p_\mathrm{s}(m, c \mid \mass, \Theta) = \mathcal{N}(m \mid m'\!,\sigma_{m})\mathcal{N}(c \mid c'\!,\sigma_{c}),
\end{equation}
where $\mathcal{N}$ is the Gaussian distribution and $(m',c')$ is the true location determined by $\mass$.
In the same way, we can derive the observed location of a binary star, $p_\mathrm{b}(m, c \mid \mass_1, q, \Theta)$, as a function of the mass of its major component $\mass_1$ and mass ratio $q$.
Together with the mass function $\mathcal{F}_\mathrm{mf}$ and binary mass-ratio distribution $\mathcal{F}_q$, we have
\begin{align}\label{eq:phi-s}
\phi_\mathrm{s}(m,c \mid \Theta) &= \!\!\int\!\! p_{\rm s}(m, c \mid \mass)\mathcal{F}_\mathrm{mf}(\mass) d\mass, \\
\phi_\mathrm{b}(m,c \mid \Theta) &= \!\!\int\!\! p_{\rm b}(m, c \mid \mass,q)\mathcal{F}_\mathrm{mf}(\mass) \mathcal{F}_q(q) d\mass dq,
\end{align}
where we omitted the condition on $\Theta$ from $p_{\rm s,b}$ and $\mathcal{F}_{\mathrm{mf},q}$ for brevity.
Both the stellar mass function and binary mass ratio are assumed to follow power-law distributions,
\begin{align}\label{eq:F-MF}
\mathcal{F}_\mathrm{mf} (\mass \mid \Theta) &= {dN}/{d\mass} \propto \mass^{\amf},\\
\mathcal{F}_q (q \mid \Theta) &= {dN}/{dq} \propto q^{\gq},
\end{align}
characterized by parameters $\amf$ and $\gq$ respectively.
For a star selected by flux limits $m_{\min}<m<m_{\max}$, 
the distributions are normalized such that
$\int_{m_{\min}}^{m_{\max}}\!\!\int_c\! \phi_\mathrm{s,b}(m,c\mid \Theta) dm dc =1$.
It is noteworthy that $\phi_\mathrm{s}$ (or $\phi_\mathrm{b}$) can be different for different stars based on their observational errors and selection function.

Specifically, we use the PARSEC theoretical isochrones \citep{bressanPARSECStellarTracks2012}\footnote{PARSEC version 1.2S, \url{http://stev.oapd.inaf.it/cgi-bin/cmd}} with the Gaia EDR3 photometric system (\citealt{rielloGaiaEarlyData2021})
and the variable extinction model YBC \citep{chenYBCStellarBolometric2019}\footnote{\url{http://stev.oapd.inaf.it/YBC}}.
For a theoretical isochrone characterized by age and metallicity,
we convert the absolute magnitude and the intrinsic color $(G, \BPRP)$ to the apparent magnitude and the reddened color $(m, c)$ according to the distance module and dust extinction.
The magnitudes of an unresolved binary are computed for the total flux,
$m = -2.5\log_{10} (10^{-0.4m_1} + 10^{-0.4 m_2} )$.

\subsection{Model of field stars}
\label{sec:field model}

Assuming that the field stars in a cluster region belong to the same population as those in its neighborhood,
we construct a nonparametric empirical model of $\phi_\mathrm{fs}(m,c)$ from an auxiliary sample of neighboring field stars,%
\footnote{
    The field sample can be stars from the neighboring sky area of the cluster, or alternatively, stars in the same sky area but distinguished from the cluster kinematically (\citealt{liMiMOMixtureModel2022}).
}
$\{m_k, c_k\}_{k=1,\ldots,N_\mathrm{fs}}$, for each cluster,
though kernel density estimation technique,
\begin{align}\label{eq:field stars_prob}
    \phi_\mathrm{fs}(m,c) {=} \frac{1}{N_\mathrm{fs}} \sum\nolimits_{k=1}^{N_\mathrm{fs}} \mathcal{N}(m \mid m_k,\epsilon_{m,k})\mathcal{N}(c \mid c_k, \epsilon_{c,k}),
\end{align}
where $\epsilon_{m,k}$ and $\epsilon_{c,k}$ are the Gaussian kernel sizes of the $k$-th star. The smoothing sizes are assigned adaptively \citep{liVersatileAccurateMethod2019} so that a star with low local density or larger observational uncertainty in the CMD has greater smoothing kernels.

\bibliography{main}{}
\bibliographystyle{aasjournal}

\end{CJK*}
\end{document}